\documentclass[3p,twocolumn]{elsarticle}
\usepackage[T1]{fontenc}
\usepackage[utf8]{inputenc}
\usepackage{lmodern}
\usepackage{epsfig} 
\usepackage{graphicx}
\usepackage{amsmath}
\usepackage{ulem}
\usepackage{color}
\usepackage{slashed}
\usepackage{subfig}
\usepackage{mathtools}
\usepackage{natbib}

\newcommand{\B}{{\scriptscriptstyle B}}

\newcommand{\cut}{\nonumber\\}
\newcommand{\avg}[1]{\left\langle #1 \right\rangle}

\newcommand{\scdots}{\cdot\hspace{-3px}\cdot\hspace{-3px}\cdot}
\DeclareMathOperator*{\SumInt}{%
	\mathchoice%
	{\ooalign{$\displaystyle\sum$\cr\hidewidth$\displaystyle\int$\hidewidth\cr}}
	{\ooalign{\raisebox{.14\height}{\scalebox{.7}{$\textstyle\sum$}}\cr\hidewidth$\textstyle\int$\hidewidth\cr}}
	{\ooalign{\raisebox{.2\height}{\scalebox{.6}{$\scriptstyle\sum$}}\cr$\scriptstyle\int$\cr}}
	{\ooalign{\raisebox{.2\height}{\scalebox{.6}{$\scriptstyle\sum$}}\cr$\scriptstyle\int$\cr}}
}
\newcommand{\DoubleBubble}{{
		\raisebox{-1.2ex}{\includegraphics[height=3.5ex]{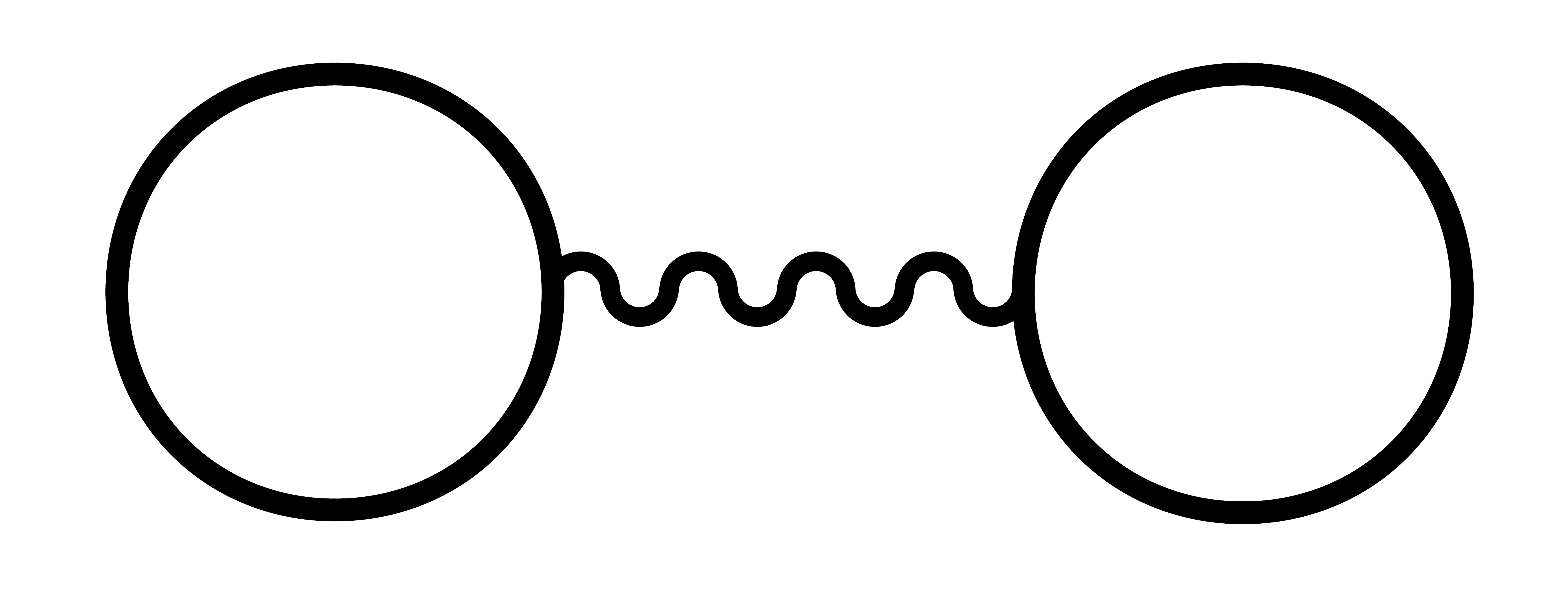}}
}}
\newcommand{\Bubble}{{
		\raisebox{-1.2ex}{\includegraphics[height=4ex]{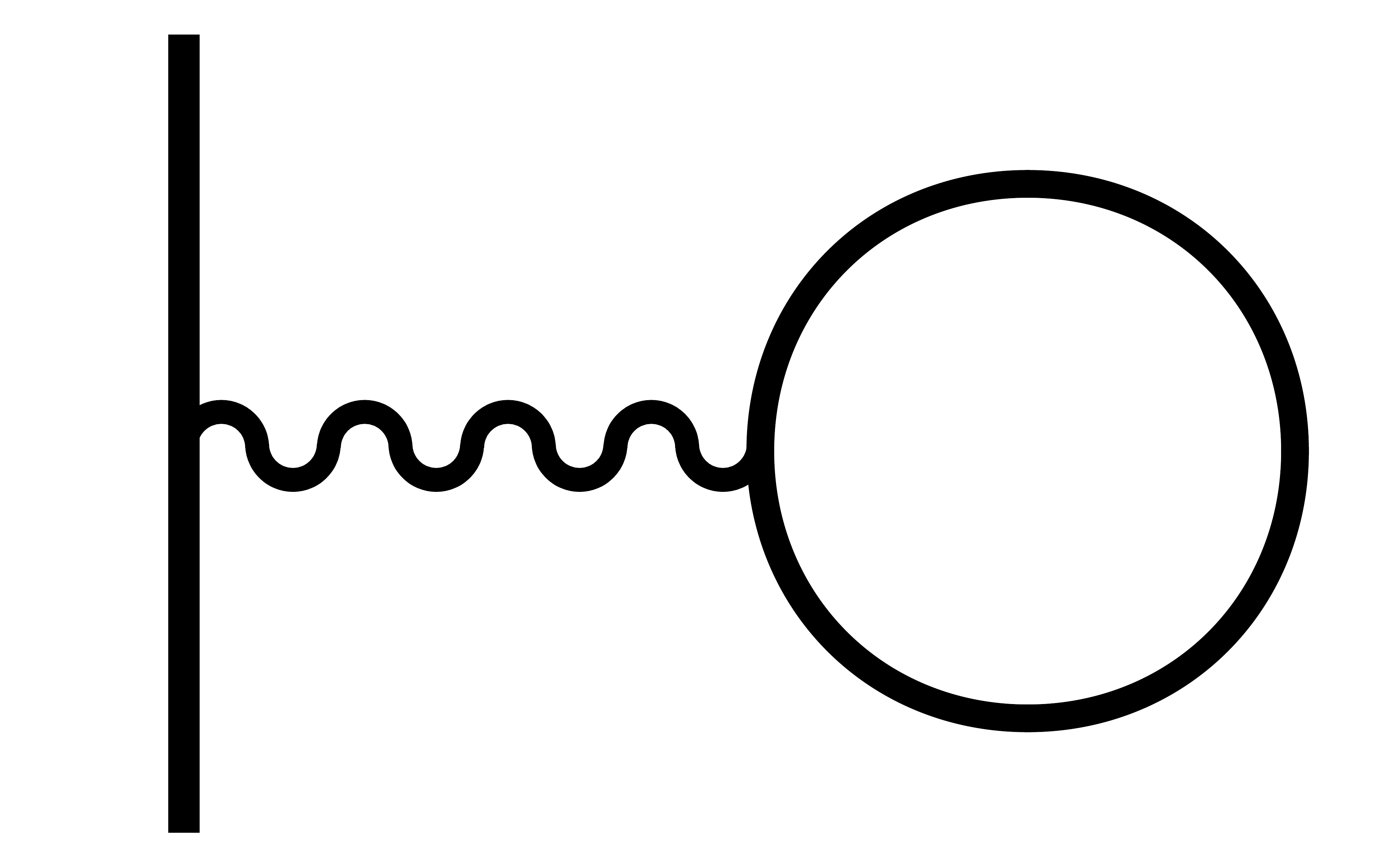}}
}}
\newcommand{\Oyster}{
	\raisebox{-1.2ex}{\includegraphics[height=3.5ex]{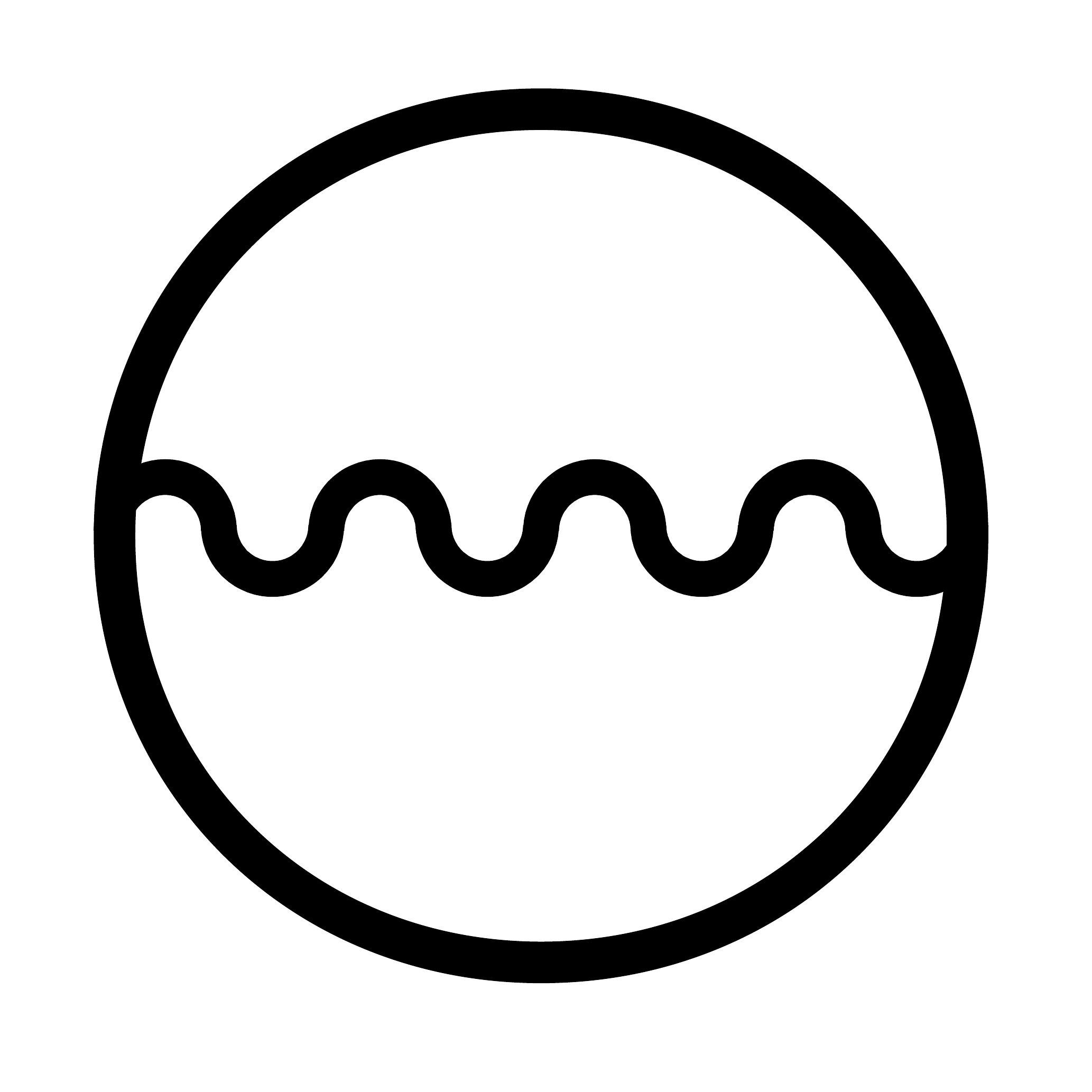}}
}

\title{Do Delta Baryons Play a Role in Neutron Stars?}
\author[1]{T. F. Motta}
\address[1]{CSSM and ARC, Department of Physics, University of Adelaide SA 5005 Australia}
\author[1]{A. W. Thomas}
\author[2]{P.~A.~M.~Guichon}
\address[2]{IRFU-CEA, Universit\'{e} Paris-Saclay, F91191 Gif sur Yvette, France}

\begin{document}
	
\begin{abstract}
The presence of exotic hadrons, such as hyperons and $\Delta$ isobars, in the dense nuclear matter in their cores has been shown to produce important changes in the properties of neutron stars. Within the quark-meson coupling model, we show that the many-body forces generated by the change in the internal quark structure of the baryons in the strong scalar mean fields generated in dense nuclear matter prohibit the appearance of $\Delta$ isobars. 
\end{abstract}

\maketitle

\section{Introduction}
The study of nuclear matter in $\beta$-equilibrium at densities above that of normal nuclear matter presents a number of outstanding challenges for modern nuclear theory. Neutron stars (NS), in  particular, constitute a remarkable laboratory for studying such effects. The predictions for the structure of these compact stars change dramatically with different assumptions about which baryons play a role. Especially in the era of gravitational wave astronomy, experimental constraints such as those provided by 
GW170817~\cite{TheLIGOScientific:2017qsa} and \cite{Demorest2010,Antoniadis2013,Arzoumanian} will definitely shed new light on this issue.

There has long been a debate about the role of hyperons in heavy 
NS~\cite{Glendenning:1982nc,Glendenning:1984}. At the time of the discovery of 
PSR J1614-2230~\cite{Demorest2010} it was widely believed that the appearance of hyperons would soften the equation of state (EoS) to such an extent that stars with masses as large as $2 M_\odot$ would not be possible, a notable exception being the work based upon the quark-meson coupling (QMC)  model~\cite{RikovskaStone:2006ta,Stone:2010jt}. Now many authors incorporate the effects of many-body forces which lead to EoS including hyperons which are compatible with the existence of heavy stars. 

There is currently a similar but as yet unresolved debate about the presence, or lack thereof, of $\Delta$ isobars in the composition of NS. We recall that at sufficiently high density the Pauli Exclusion Principle ensures the stability of a $\Delta$ baryon once the conditions of chemical equilibrium permit its appearance. Early results~\cite{Glendenning:1984,Glendenning:1997wn,Glendenning:1991es} showed very high critical densities for the appearance of $\Delta$ baryons, as high as $9\rho_0$ (with $\rho_0$ the saturation density of symmetric nuclear matter). However, with better constraints on the properties of nuclear matter at saturation, such as the symmetry energy, and further developments in models for infinite nuclear matter, many different predictions have appeared~\cite{Li:2019,Li:2018,Cai:2015,Drago:2014oja,Chen:2009am,Xiang:2003qz,Huber:1998mj}, which typically show much lower values for the critical density at which the $\Delta^-$ should appear. The $\Delta^-$ is, of course, the isobar which appears first because its chemical potential only need equal that of the neutron minus that of the electron, as explained later.

The consequences for NS properties of the appearance of a $\Delta^-$ component have been shown to be very significant. For example, it has been shown that the tidal deformability (TD) for a typical NS of mass $1.4 M_\odot$ could be reduced by as much as 300 for a reasonably 
attractive $\Delta$-nucleus potential~\cite{Li:2019}. Furthermore, Li and co-workers~\cite{Li:2018} have shown that the softening of the EoS caused by the appearance of $\Delta$ baryons could reduce the radius of a star of mass $1.4 M_\odot$ by as much as 2km. These are dramatic effects which may well be tested against data from gravitational wave detection~\cite{TheLIGOScientific:2017qsa} and the NICER mission~\cite{Gendreau2016} in the near future. It is therefore crucial to explore the physics underlying the interactions of the $\Delta$ in dense nuclear matter.

The quark-meson coupling (QMC) model~\cite{Guichon:1987jp,Guichon:1995ue,Saito:2005rv,review18} starts with a quark model for the structure of baryons (for example, the MIT bag model~\cite{Bag}) and then solves self-consistently for the changes in structure induced by the strong scalar fields which are known to occur in dense nuclear matter. At the baryon level in nuclear matter the change in hadron structure in-medium leads to a suppression of the baryon coupling to the mean scalar field as the density rises. This is often written in terms of a scalar polarizability. At the level of the energy density functional this can be shown to be equivalent to including repulsive three-body forces between all baryons with strengths {\it predicted} by the model {\it with no additional parameters}~\cite{Guichon:2004xg,06}. 

With regard to the binding of hyperons in nuclei, a crucial development was the realization that the so-called ``hyperfine'' interaction between quarks, arising from the spin-dependent one-gluon-exchange (OGE) interaction, must be enhanced in-medium~\cite{Guichon:2008zz}. Essentially, in the bag model the strength of the hyperfine interaction involves the inverse of the eigenenergies of the two interacting quarks. Hence, in the QMC model, where the scalar field does not couple to the strange quark, the spin dependent interaction between two light quarks grows considerably faster with density in-medium than that between one strange quark and one non-strange quark. As the mass difference between a $\Sigma$ and $\Lambda$ baryon arises primarily from the hyperfine interaction~\cite{Bag,Thomas:1982kv}, the $\Sigma-\Lambda$ mass splitting grows with density. This provides a very natural explanation~\cite{Tsushima:2009zh,Tsushima:2010ew} of the observed repulsive $\Sigma$-nucleus interaction and the consequent absence of $\Sigma$-hypernuclei. 
It also leads to the complete absence 
of $\Sigma$-hyperons in NS within the QMC model~\cite{RikovskaStone:2006ta,Motta2019}.

It is this OGE hyperfine interaction which splits the $\Delta$ from the nucleon in free space in essentially all quark models and, as we have just explained, this mass difference will be enhanced in-medium. Indeed, already at twice nuclear matter density the mass splitting is enhanced by almost 100 MeV. This already suggests that within the QMC model it is unlikely that the $\Delta$  baryon will make an appearance. However, a firm conclusion requires a detailed study including the effect of the iso-vector scalar interaction which will naturally tend to favour the 
$\Delta^-$ in matter that contains predominantly neutrons.

Our aim is to study the chemical potential of the $\Delta$ in matter in $\beta$-equilibrium within the QMC model in order to understand whether it is likely to appear at any density relevant to the physics of neutron stars. In Section~\ref{modelind} we review the derivation of the EoS of dense matter in the model, along with the condition for the $\Delta^-$ to appear under the constraints of $\beta$-equilibrium. In Section~\ref{qmcsec}, we present the numerical results, showing explicitly that the $\Delta$ baryons do not play a role within the QMC model. Finally, section~\ref{sumcon} is devoted to a brief summary and discussion.

\section{Equation of State for Dense Nuclear Matter}\label{modelind}

The QMC model~\cite{94,Stone:2010jt,review18} is based on a quark-level description of the baryons as bags of three confined quarks that couple directly to meson fields:
\begin{align}
\label{eq:1}
	\mathcal{L}_\text{quarks}=&\bar\psi_q(i\slashed\partial - m_q)\psi_q  + \mathcal{L}_{\text{I}} + \mathcal{B}\cut
	\mathcal{L}_{\text{I}}=& \bar\psi_q(g_\sigma^q\hat{\sigma} + g_\omega^q{\hat{\slashed\omega}} + g_\rho^q{\hat{{\boldsymbol{\slashed b}}}\cdot\boldsymbol{\tau}} 
	+ g_\delta^q{\hat{{\boldsymbol{\delta}}}\cdot\boldsymbol{\tau}})\psi_q \, ,
\end{align}
where we denote the isoscalar scalar and vector fields as $\sigma$ and $\omega$ and the corresponding isovector fields as $\boldsymbol{\delta}$ and $\boldsymbol{b}$ (with the $\rho$ meson field 
denoted $\boldsymbol{b}$ to avoid confusion with the density).
This problem is solved for a spherically symmetric system of confined quarks subject to the standard linear and non-linear boundary conditions. The former, $(1+i\vec{\gamma}\cdot\hat{r})\psi_q=0$ at $r=R_B$, simulates confinement and the latter ensures stability. At the level of baryons, the underlying quark structure is manifest through the field dependent couplings to the scalar fields:
\begin{align}
\label{baryonlag}
	\mathcal{L}_\text{QMC}=&\bar\Psi(i\slashed\partial - M_\B)\Psi + \bar\Psi\big(g_\sigma(\sigma,\delta)\hat{\sigma} + g_\omega{\hat{\slashed\omega}} \cut 
	&+ g_\rho{\hat{{\boldsymbol{\slashed b}}}\cdot\boldsymbol{\tau}} 
	+ g_\delta(\sigma,\delta){\hat{{\boldsymbol{\delta}}}\cdot\boldsymbol{\tau}}\big)\Psi \, .
\end{align}

The solution of the quark level bag equations coupled to meson fields yields an expression for the rest energy of the baryon that depends non trivially on the meson fields
\begin{align}
	M^\star_\B(\sigma,\delta)=\sum_f \frac{n_f\Omega_f-z_0}{R_\B} + \mathcal{B}V_\B + \Delta_{EM} \, ,
\end{align}
where $n_f$ is the number of quarks of flavour $f$ in the bag, $\Omega_f$ is the corresponding eigenvalue of the Dirac equation, $\mathcal{B}$ the bag constant, $z_0$ the zero point correction, $V_B$ the volume of the bag, and $\Delta_{EM}$ the OGE hyperfine colour interaction~\cite{DeGrand1975}. The parameters $\mathcal{B}$, $z_0$ and the colour hyperfine constant 
$\alpha_c$, appearing in $\Delta_{EM}$, are chosen to reproduce the mass spectrum of the baryon octet in free space.

We fit the result of the calculation of the baryon mass obtained by solving the bag model over a wide range of values of the applied scalar fields to obtain the functional dependence of the effective baryon mass on the meson fields
\begin{eqnarray}
\label{effmassfit}
M_\B^\star(\sigma,\delta)&=&M_\B - w_{1}g_\sigma \sigma -w_{2} g_\delta  \delta + w_{3}\frac{\sigma^2}{2} \nonumber \\
&& +  w_{4}\frac{\delta^2}{2} +  w_5{\sigma\delta} \, .
\end{eqnarray}
This procedure defines the system described by the Lagrangian density in Eq.~\ref{baryonlag} and we can then proceed to solve for the energy density in Hartree-Fock approximation\footnote{The full expression for the Hartree-Fock energy density in the QMC model can be found in \ref{ApA}.}, from which we can obtain the chemical potentials  and pressure:
\begin{align}
\label{manybody}
\epsilon(n_1,...n_N) =& \sum_i \avg{\mathcal{K}} +\sum_{i,j} \Big( \DoubleBubble + \Oyster + \scdots \Big) \cut
\mu_i =& \frac{\partial \epsilon}{\partial n_i} \cut
P =& \sum_f n_f \mu_f - \epsilon \, .
\end{align}
%

\subsection{Equilibrium Conditions}
In the case of NS, where the time scales for weak interactions which change strangeness by up to one unit are much shorter than the formation time, we must minimise the energy density for the system 
in $\beta$-equilibrium, subject to the conservation of baryon number and electric charge. That is, we must minimize the function
\begin{align*}
	\epsilon(n_1,...n_N)
	-\lambda_1(n_\B-\sum_{f}n_f^\B )
	-\lambda_2(\sum Q_f n_f) \, ,
\end{align*}
where $n_f^\B$ denotes the density of a baryon of flavour $f$, $n_f$ either a baryon or a lepton of flavour $f$, and $Q_f$ the charge of the particle. The parameters $\lambda_{1,2}$ are Lagrange multipliers.
By minimizing this function for each value of the total baryon density, $n_\B$, we obtain the energy density (and consequently, through Eq.~\ref{manybody}, the pressure and chemical potentials) as a function of the total baryon density, $\rho_\B$.

\subsection{Creation Condition for the $\boldsymbol{\Delta^-}$}

Provided that the density of neutrons is sufficiently high, Pauli blocking ensures the stability of 
a $\Delta^-$ baryon once it is formed. In free space the decay of the $\Delta^-$ at rest leads to a neutron of momentum of order $k \approx 220$ MeV. In pure neutron matter the reaction would be Pauli blocked at a density $k^3/(3 \pi)^2 \approx 0.05$ fm$^{-3}$. Though this is merely a back of the envelope estimate, it is clear that at the densities typical of the neutron star core the $\Delta^-$ may be treated as a stable particle. It can be produced in reactions such as 
$$\Delta^- \rightarrow n + e^- + \nu_e$$
and 
$$ \Delta^- + p\rightarrow n + n \, $$
and its presence in $\beta$-equilibrium with the other components requires that the chemical potentials satisfy the relation:
\begin{align}
\beta\text{-equilibrium condition:}&\quad \mu_{\Delta^-} = \mu_n + \mu_e \, ,
\end{align}
provided that the neutrinos are not trapped in the star.
\begin{figure}
	\centering \includegraphics[width=0.9\linewidth]{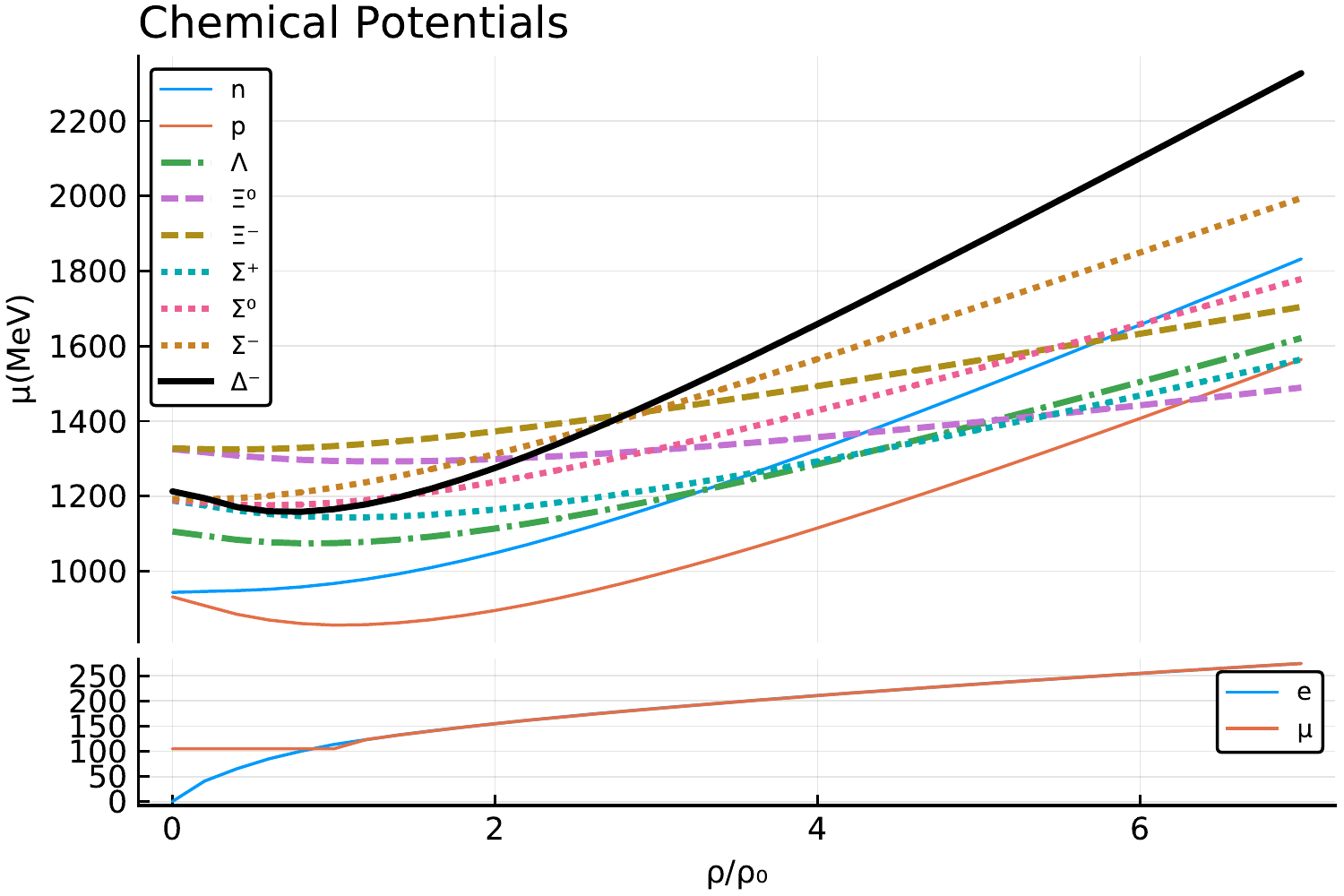}
	\caption{Chemical potentials for all baryons calculated through the QMC model.}
	\label{fig:chempots}
\end{figure}

As our focus here is whether or not the $\Delta$ can actually appear under the conditions of $\beta$-equilibrium, we need only evaluate the chemical potential for a single $\Delta^-$ baryon at rest. 
The relevant condition is, then
\begin{align}
	M_{\Delta^-} +  \sum_{\varphi,\B} \Bubble_{\hspace{-0.7cm}\varphi}^{\hspace{-5pt}\B} = \mu_n + \mu_e \, ,
\end{align}
or explicitly
\begin{align}\label{cond}
M_{\Delta^-} -  \sum_{\varphi,\B} g_{\Delta\varphi}{\bar\varphi(\rho_\B)} = \mu_n + \mu_e \, .
\end{align}
where $\bar\varphi$ are the mean field values for the mesons and the scalar couplings are evaluated at the 
relevant density.

Of course, even though there are no Fock terms involving $\Delta$ baryons, if one takes seriously the chiral structure of the baryons~\cite{Thomas:1982kv,Theberge:1980ye,Thomas:1981vc,Miyatsu:2019xub} there is a potentially significant correction to the chemical potential of the $\Delta$ arising from Pauli blocking of the intermediate nucleon in the process $\Delta \rightarrow N + \pi$. The corresponding effect on the nucleon self-energy is already taken into account through the Fock term associated with pion exchange. As the latter is less than 20MeV up to densities 6 times that of nuclear matter~\cite{danielcarroll}, we expect the corresponding effect for the $\Delta$ to be of a similar magnitude. Indeed, explicit calculation confirms that the effect of Pauli blocking on the mass of the $\Delta$ only exceeds 10 MeV when the Fermi momentum is close to the momentum of the momentum of the neutron resulting from $\Delta^-$ decay at rest, and it is well below 10MeV above  nuclear matter density. As we shall see below, such small shifts have no effect on our conclusion.

\section{Numerical results}\label{qmcsec}

As in earlier work~\cite{Motta2019}, the coupling constants of the $\sigma, \omega$ and 
$\rho$ mesons in free space have been chosen (with the value of the $\delta$ meson coupling taken from Ref.~\cite{Haidenbauer}) to reproduce the properties of nuclear matter at the saturation density, $\rho_0=0.16$ fm$^{-3}$, namely the binding energy per nucleon, $\varepsilon= 15.8$ MeV, the symmetry energy, $S=30$ MeV, and the slope of the symmetry energy $L_0 = 63$ MeV.   As the coupling constant of the $\delta$ meson is less well constrained, we have chosen to test the 
dependence of our conclusions on this choice by repeating the calculations with this coupling ($G_\delta = g_\delta^2/m_\delta^2$) either set to zero or doubled, while retaining the same nuclear matter properties.
\begin{figure}
	\centering \includegraphics[width=0.9\linewidth]{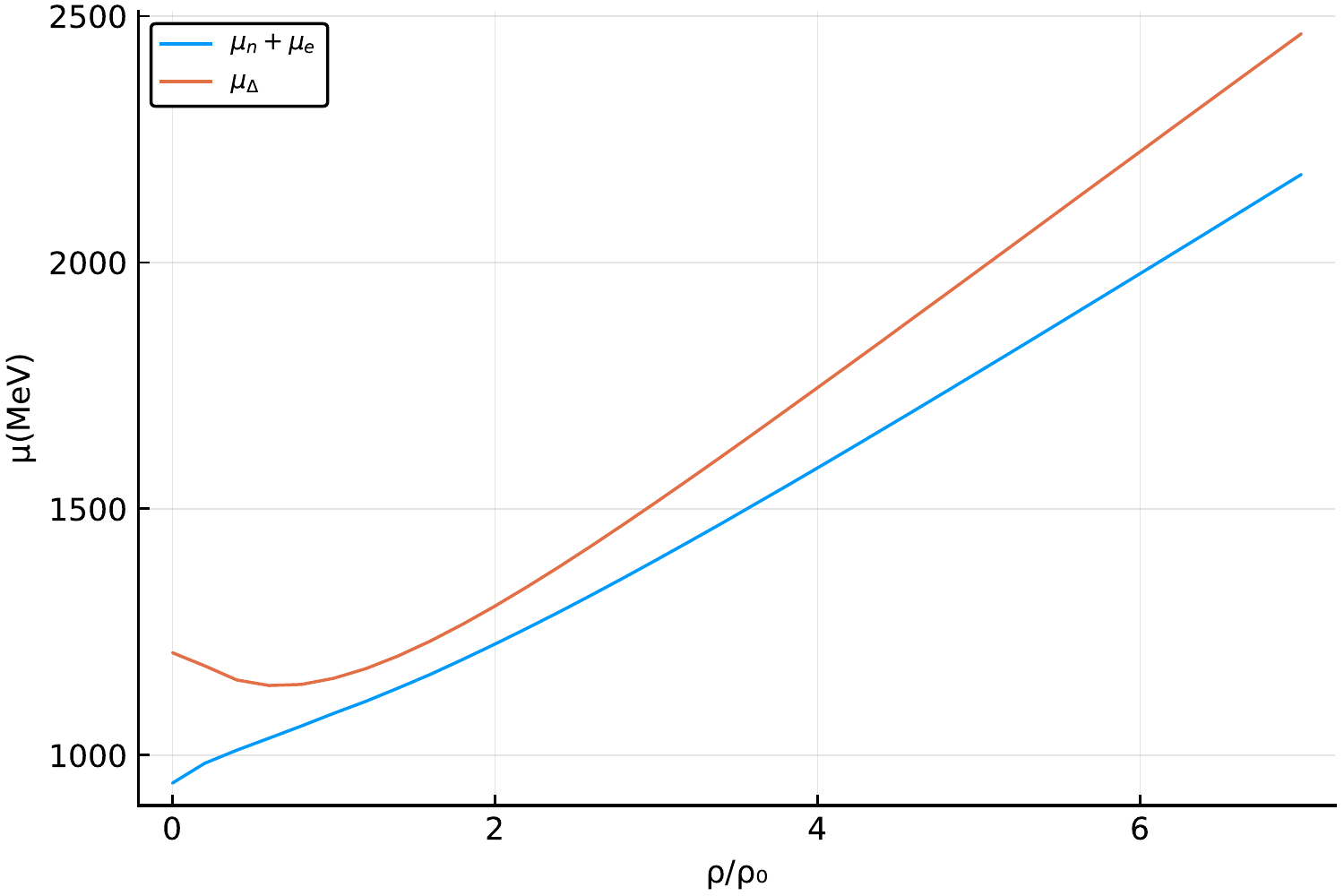}
	\caption{Condition for the appearance of the $\Delta^-$ under $\beta$-equilibrium.}
	\label{fig:deltachempots}
\end{figure}
We stress that in the QMC model the fundamental couplings are to the light quarks and while the fitting is done at the nucleon level those values determine the meson-quark couplings which, in turn, determine the couplings to {\it all other baryons}, with no further parameters.

In Figure \ref{fig:chempots} we show the chemical potentials calculated within the model (with couplings $G_\sigma=10.0$fm$^2$, $G_\delta=3.0$fm$^2$, $G_\omega=6.4$fm$^2$, $G_\rho=4.27$fm$^2$). 
Figure~\ref{fig:deltachempots} shows that the value of the chemical potential for the $\Delta^-$ not only lies above threshold for all values of the baryon density but the difference $\mu_{\Delta^-} - (\mu_n+\mu_e)$ actually grows with the value of $\rho_\B$.
It then becomes clear that there is no value of the density at which the equality in Eq.~\ref{cond} holds and the value for $\mu_n + \mu_e$ is always smaller than $\mu_{\Delta^-}$. In other words, the $\Delta$ baryon is always too ``expensive'' to produce. 

\begin{figure}
	\centering \includegraphics[width=0.9\linewidth]{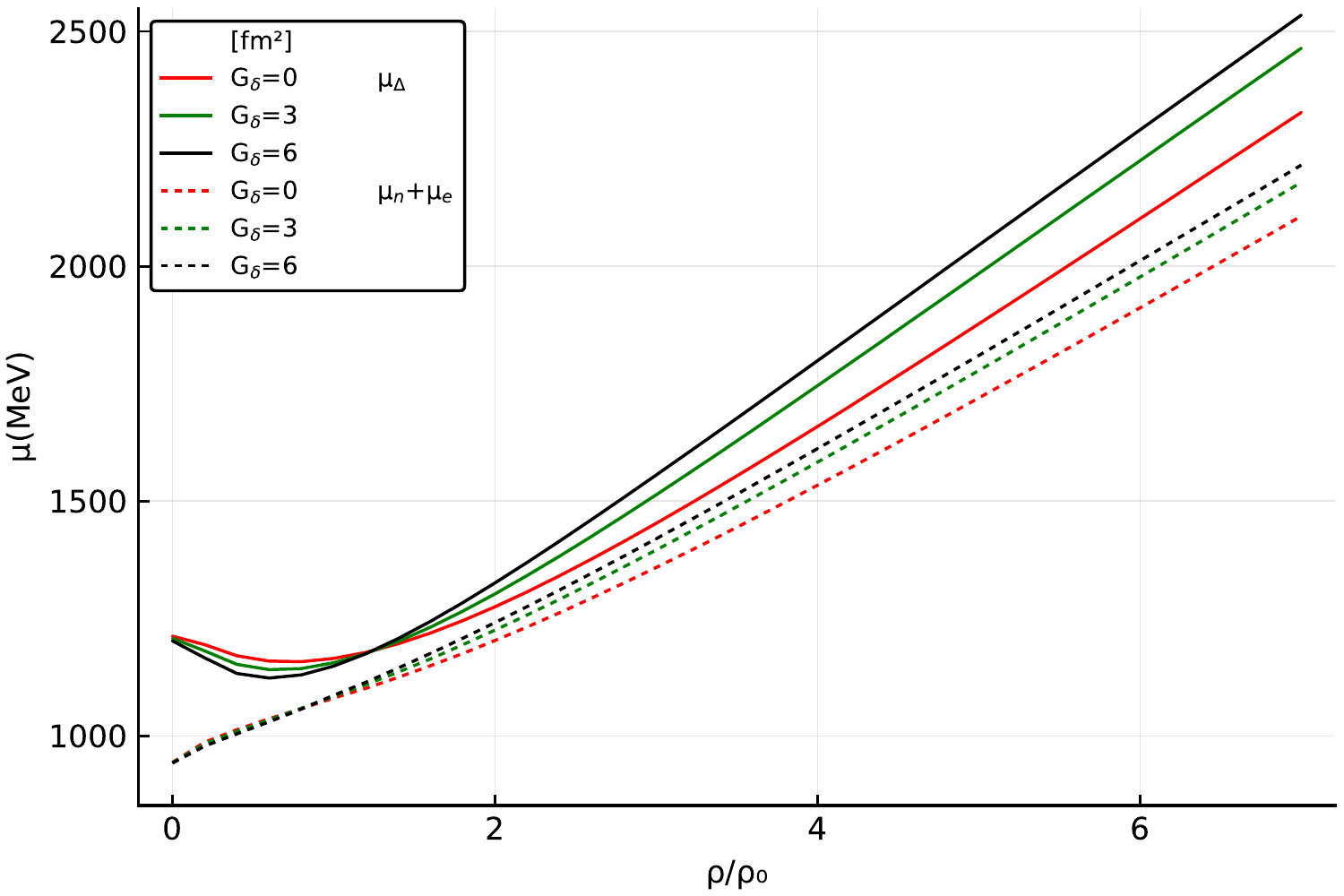}
	\caption{Comparison of $\mu_\Delta$ for different choices of the coupling of the isovector scalar 
meson, $\delta$, to the nucleon.}
	\label{fig:compareDelta}
\end{figure}

Finally the QMC model naturally creates a density dependent coupling of the baryons with the scalar sector. This implies that the isovector-scalar meson also has a density dependent coupling. It is therefore important to examine the variation in the $\Delta^-$ chemical potential for the two alternative choices of $G_\delta$. As we see in Fig.~\ref{fig:compareDelta}, even over this very wide range there is no value such that the $\Delta^-$ can actually satisfy the condition necessary to appear in the NS.
%
\section{Summary and Conclusions}\label{sumcon}

We have shown that the repulsive many-body forces that arise naturally in the QMC model from the quark substructure of the baryons increases the chemical potential of the $\Delta^-$ in such a way that it simply cannot appear in a neutron star. Given that this conclusion differs from the findings of most other studies, many of which involve considerably more parameters, it is worthwhile to review the physics behind it. In the QMC model, the change of the internal structure of a baryon in-medium plays a crucial role. For symmetric nuclear matter containing nucleons the internal response to the scalar mean field reduces the effective coupling to that field as the density rises, providing a novel saturation mechanism. As we explained in detail, this change of internal structure also enhances the hyperfine interaction between non-strange quarks, which is repulsive in the $\Sigma$ and attractive in the $\Lambda$ hyperon. This naturally leads to an effective potential for the $\Sigma$ which is repulsive, explaining the absence of $\Sigma$-hypernuclei and predicting that $\Sigma$ hyperons should not appear in NS. Exactly the same effect leads to a large enhancement of the mass splitting between the $\Delta$ and the nucleon in dense matter and this is what forbids the appearance of the $\Delta^-$ in NS. At the baryon level this effect can be described as a particularly repulsive $\Delta - N - N$ three-body force but the key point is that within the QMC model this is predicted with no additional parameters.

Future measurements of gravitational waves and the upcoming NICER experiment \cite{NICER}, which intends to measure the radius of a neutron star with great precision, as well as future terrestrial experiments may shed more light on this question. 

\section*{Acknowledgements}
This work was supported by the University of Adelaide and the Australian Research Council through Discovery Projects DP150103101 and DP180100497.

\bibliographystyle{ieeetr}
\bibliography{bibliography}

\begin{thebibliography}{10}

\bibitem{TheLIGOScientific:2017qsa}
B.~Abbott {\em et~al.}, ``{GW170817: Observation of Gravitational Waves from a
  Binary Neutron Star Inspiral},'' {\em Phys. Rev. Lett.}, vol.~119, no.~16,
  p.~161101, 2017.

\bibitem{Demorest2010}
P.~Demorest, T.~Pennucci, S.~Ransom, M.~Roberts, and J.~Hessels, ``{Shapiro
  Delay Measurement of A Two Solar Mass Neutron Star},'' {\em Nature},
  vol.~467, pp.~1081--1083, 2010.

\bibitem{Antoniadis2013}
J.~Antoniadis {\em et~al.}, ``{A Massive Pulsar in a Compact Relativistic
  Binary},'' {\em Science}, vol.~340, p.~6131, 2013.

\bibitem{Arzoumanian}
Z.~Arzoumanian {\em et~al.}, ``{The NANOGrav 11-year Data Set: High-precision
  timing of 45 Millisecond Pulsars},'' {\em Astrophys. J. Suppl.}, vol.~235,
  no.~2, p.~37, 2018.

\bibitem{Glendenning:1982nc}
N.~K. Glendenning, ``{THE HYPERON COMPOSITION OF NEUTRON STARS},'' {\em Phys.
  Lett.}, vol.~114B, pp.~392--396, 1982.

\bibitem{Glendenning:1984}
N.~K. Glendenning, ``{Neutron Stars Are Giant Hypernuclei?},'' {\em Astrophys.
  J.}, vol.~293, pp.~470--493, 1985.

\bibitem{RikovskaStone:2006ta}
J.~Rikovska-Stone, P.~A.~M. Guichon, H.~H. Matevosyan, and A.~W. Thomas,
  ``{Cold uniform matter and neutron stars in the quark-mesons-coupling
  model},'' {\em Nucl. Phys.}, vol.~A792, pp.~341--369, 2007.

\bibitem{Stone:2010jt}
J.~R. Stone, P.~A.~M. Guichon, and A.~W. Thomas, ``{Role of Hyperons in Neutron
  Stars},'' 2010.

\bibitem{Glendenning:1997wn}
N.~K. Glendenning, ``{Compact stars: Nuclear physics, particle physics, and
  general relativity, 2nd ed},'' {\em New York, USA: Springer (2000) ISBN
  978-0-387-98977-8}.

\bibitem{Glendenning:1991es}
N.~K. Glendenning and S.~A. Moszkowski, ``{Reconciliation of neutron star
  masses and binding of the lambda in hypernuclei},'' {\em Phys. Rev. Lett.},
  vol.~67, pp.~2414--2417, 1991.

\bibitem{Li:2019}
J.~J. Li and A.~Sedrakian, ``{Implications from GW170817 for $\Delta$-isobar
  admixed hypernuclear compact stars},'' {\em Astrophys. J. Lett.}, vol.~874,
  p.~L22, 2019.

\bibitem{Li:2018}
J.~J. Li, A.~Sedrakian, and F.~Weber, ``{Competition between delta isobars and
  hyperons and properties of compact stars},'' {\em Phys. Lett.}, vol.~B783,
  pp.~234--240, 2018.

\bibitem{Cai:2015}
B.-J. Cai, F.~J. Fattoyev, B.-A. Li, and W.~G. Newton, ``{Critical density and
  impact of $\Delta$(1232) resonance formation in neutron stars},'' {\em Phys.
  Rev.}, vol.~C92, no.~1, p.~015802, 2015.

\bibitem{Drago:2014oja}
A.~Drago, A.~Lavagno, G.~Pagliara, and D.~Pigato, ``{Early appearance of Delta
  isobars in neutron stars},'' {\em Phys. Rev.}, vol.~C90, no.~6, p.~065809,
  2014.

\bibitem{Chen:2009am}
Y.~Chen, Y.~Yuan, and Y.~Liu, ``{Neutrino mean free path in neutron star matter
  with Delta isobars},'' {\em Phys. Rev.}, vol.~C79, p.~055802, 2009.

\bibitem{Xiang:2003qz}
H.~Xiang and G.~Hua, ``{Delta excitation and its influences on neutron stars in
  relativistic mean field theory},'' {\em Phys. Rev.}, vol.~C67, p.~038801,
  2003.

\bibitem{Huber:1998mj}
H.~Huber, F.~Weber, and M.~K. Weigel, ``{Symmetric and asymmetric nuclear
  matter in the relativistic approach at finite temperatures},'' {\em Phys.
  Rev.}, vol.~C57, pp.~3484--3487, 1998.

\bibitem{Gendreau2016}
K.~C. Gendreau {\em et~al.}, ``The neutron star interior composition explorer
  (nicer): design and development,'' 2016.

\bibitem{Guichon:1987jp}
P.~A.~M. Guichon, ``{A Possible Quark Mechanism for the Saturation of Nuclear
  Matter},'' {\em Phys. Lett.}, vol.~B200, pp.~235--240, 1988.

\bibitem{Guichon:1995ue}
P.~A.~M. Guichon, K.~Saito, E.~N. Rodionov, and A.~W. Thomas, ``{The Role of
  nucleon structure in finite nuclei},'' {\em Nucl. Phys.}, vol.~A601,
  pp.~349--379, 1996.

\bibitem{Saito:2005rv}
K.~Saito, K.~Tsushima, and A.~W. Thomas, ``{Nucleon and hadron structure
  changes in the nuclear medium and impact on observables},'' {\em Prog. Part.
  Nucl. Phys.}, vol.~58, pp.~1--167, 2007.

\bibitem{review18}
P.~Guichon, J.~Stone, and A.~Thomas, ``Quark–meson-coupling (qmc) model for
  finite nuclei, nuclear matter and beyond,'' {\em Progress in Particle and
  Nuclear Physics}, 2018.

\bibitem{Bag}
A.~Chodos, R.~L. Jaffe, K.~Johnson, and C.~B. Thorn, ``Baryon structure in the
  bag theory,'' {\em Phys. Rev. D}, vol.~10, pp.~2599--2604, Oct 1974.

\bibitem{Guichon:2004xg}
P.~A.~M. Guichon and A.~W. Thomas, ``{Quark structure and nuclear effective
  forces},'' {\em Phys. Rev. Lett.}, vol.~93, p.~132502, 2004.

\bibitem{06}
P.~A.~M. Guichon, H.~H. Matevosyan, N.~Sandulescu, and A.~W. Thomas,
  ``{Physical origin of density dependent force of the skyrme type within the
  quark meson coupling model},'' {\em Nucl. Phys.}, vol.~A772, pp.~1--19, 2006.

\bibitem{Guichon:2008zz}
P.~A.~M. Guichon, A.~W. Thomas, and K.~Tsushima, ``{Binding of hypernuclei in
  the latest quark-meson coupling model},'' {\em Nucl. Phys.}, vol.~A814,
  pp.~66--73, 2008.

\bibitem{Thomas:1982kv}
A.~W. Thomas, ``{Chiral Symmetry and the Bag Model: A New Starting Point for
  Nuclear Physics},'' {\em Adv. Nucl. Phys.}, vol.~13, pp.~1--137, 1984.

\bibitem{Tsushima:2009zh}
K.~Tsushima, P.~A.~M. Guichon, R.~Shyam, and A.~W. Thomas, ``{Binding of
  hypernuclei, and phtoproduction of Lambda-hypernuclei in the latest
  quark-meson coupling model},'' {\em Int. J. Mod. Phys.}, vol.~E19,
  pp.~2546--2551, 2010.
\newblock [,254(2009)].

\bibitem{Tsushima:2010ew}
K.~Tsushima and P.~A.~M. Guichon, ``{Hypernuclei in the quark-meson coupling
  model},'' {\em AIP Conf. Proc.}, vol.~1261, no.~1, pp.~232--237, 2010.

\bibitem{Motta2019}
T.~F. Motta, A.~M. Kalaitzis, S.~Antić, P.~A.~M. Guichon, J.~R. Stone, and
  A.~W. Thomas, ``{Isovector Effects in Neutron Stars, Radii and the GW170817
  Constraint},'' 2019.

\bibitem{94}
K.~Saito and A.~W. Thomas, ``{A Quark - meson coupling model for nuclear and
  neutron matter},'' {\em Phys. Lett.}, vol.~B327, pp.~9--16, 1994.

\bibitem{DeGrand1975}
T.~A. DeGrand, R.~L. Jaffe, K.~Johnson, and J.~E. Kiskis, ``{Masses and Other
  Parameters of the Light Hadrons},'' {\em Phys. Rev.}, vol.~D12, p.~2060,
  1975.

\bibitem{Theberge:1980ye}
S.~Theberge, A.~W. Thomas, and G.~A. Miller, ``{The Cloudy Bag Model. 1. The
  (3,3) Resonance},'' {\em Phys. Rev.}, vol.~D22, p.~2838, 1980.
\newblock [Erratum: Phys. Rev.D23,2106(1981)].

\bibitem{Thomas:1981vc}
A.~W. Thomas, S.~Theberge, and G.~A. Miller, ``{The Cloudy Bag Model of the
  Nucleon},'' {\em Phys. Rev.}, vol.~D24, p.~216, 1981.

\bibitem{Miyatsu:2019xub}
T.~Miyatsu and K.~Saito, ``{Equation of State for Neutron Stars in the
  Quark-Meson Coupling Model with the Cloudy Bag},'' 2019.

\bibitem{danielcarroll}
D.~L. Whittenbury, J.~D. Carroll, A.~W. Thomas, K.~Tsushima, and J.~R. Stone,
  ``{Quark-Meson Coupling Model, Nuclear Matter Constraints and Neutron Star
  Properties},'' {\em Phys. Rev.}, vol.~C89, p.~065801, 2014.

\bibitem{Haidenbauer}
J.~Haidenbauer, K.~Holinde, and A.~W. Thomas, ``{Investigation of pion exchange
  in the N N and N anti-N systems},'' {\em Phys. Rev.}, vol.~C45, pp.~952--958,
  1992.

\bibitem{NICER}
A.~L. Watts {\em et~al.}, ``{Colloquium : Measuring the neutron star equation
  of state using x-ray timing},'' {\em Rev. Mod. Phys.}, vol.~88, no.~2,
  p.~021001, 2016.

\end{thebibliography}

\appendix
\section{HF-QMC}\label{ApA}
In Hartree-Fock approximation, the QMC model yields the following expression for energy density.

Kinetic:
\begin{align}
	\avg{\mathcal{K}}=&\frac{1}{\pi^2}\sum_B\int_{0}^{k_F^B}{k^2}{\sqrt{k^2+M_\B^\star(\sigma,\delta)^2}dk} +  
	\cut&\frac{1}{\pi^2}\sum_L \int_{0}^{k_F^L}{k^2}{\sqrt{k^2+m_L^2}dk}
\end{align}

Hatree:
\begin{eqnarray}
\epsilon_\text{H}=& \frac{m_\sigma^2\sigma^2}{2} + \frac{m_\omega^2\omega^2}{2}+ \frac{m_b^2b^2}{2} + \frac{m_\delta^2\delta^2}{2}
\end{eqnarray}
where the sum is over entire baryon octet $B=(n, p, \Lambda, \Sigma^{0,\pm},  \Xi^{0-})$ and leptons $L=(e^-, \mu^-)$. 

The mean field mesons are given by
\begin{eqnarray}\label{scc}
m_\sigma^2\sigma =&  \sum_B \big(-\partial_\sigma M^\star_\B(\sigma,\delta)\big)\times\cut&\frac{1}{\pi^2}\int_0^{k_F^\B} k^2 
\frac{M_\B^\star(\sigma,{\delta})}{\sqrt{k^2 + M_\B^\star(\sigma,{\delta})^2}}dk  \\
m_\omega^2\omega=&\sum_B  n_\B g_\omega \times \left( 1+\frac{s_\B}{3} \right)=
\sum_B  n_\B g_\omega^\B \\
m_\rho^2b=&\sum_B  n_\B g_\rho \times t_{3\B} =\sum_B  n_\B g_\rho^\B \label{meanrho} \\
m_\delta^2{\delta} =&  \sum_B \big(-\partial_\delta M^\star_\B(\sigma,\delta)\big)
\times\cut& \frac{1}{\pi^2}\int_0^{k_F^\B} k^2 
\frac{M_\B^\star(\sigma,{\delta})}{\sqrt{k^2 + M_\B^\star(\sigma,{\delta})^2}}dk \, . 
\end{eqnarray}

And Fock:
{\begin{eqnarray}
	\epsilon_\text{F} =& 
	\frac{1}{(2\pi)^6}\SumInt_{\B, k_1,k_2} 
	\frac{\partial_\sigma M^\star_\B(\sigma,\delta)^2}{(\vec{k_1}-\vec{k_2})^2+m_\sigma^2}\times \cut&
	\left[\frac{M^\star_\B(\sigma,\delta)}{\sqrt{k_1^2+M^\star_\B(\sigma,\delta)^2}}\right] 
	\left[\frac{M^\star_{\B'}(\sigma,\delta)}{\sqrt{k_2^2+M^\star_{\B'}(\sigma,\delta)^2}}\right]\nonumber \\
	& +\frac{1}{(2\pi)^6}\SumInt_{\B,\B', k_1,k_2}  \
	\frac{Z_{t_{3\B} t_{3\B'}}}{(\vec{k_1}-\vec{k_2})^2+m_\delta^2}\times\cut& 
	\left[\frac{M^\star_\B(\sigma,\delta)}{\sqrt{k_1^2+M^\star_\B(\sigma,\delta)^2}}\right] 
	\left[\frac{M^\star_{\B'}(\sigma,\delta)}{\sqrt{k_2^2+M^\star_{\B'}(\sigma,\delta)^2}}\right] \cut
	& -\frac{1}{(2\pi)^6} \SumInt_{\B, k_1,k_2}  \frac{{g^\B_\omega}^2}{(\vec k_1 - \vec k_2)^2 + m_\omega^2} \cut&
	- \SumInt_{\B,\B', k_1,k_2}  \frac{g_\rho^2 I_{t_{3\B} t_{3\B'}}}{(\vec k_1 - \vec k_2)^2 + m_\rho^2}  \, , 
	\nonumber
	\end{eqnarray}}
where
\begin{eqnarray}
I_{t_{3\B} t_{3\B'}}
=  \delta_{t_{3\B} t_{3\B'}} + {(\delta_{t_{3\B}, t_{3\B'}+1}+\delta_{t_{3\B'}, t_{3\B}+1})}{t_\B} 
\end{eqnarray}
and
{\begin{eqnarray*}
		Z_{t_{3\B} t_{3\B'}}
		= \partial_\delta M^\star_\B(\sigma,\delta)\partial_\delta M^\star_{\B'}(\sigma,\delta)\times \delta_{t_{3\B} t_{3\B'}} \cut+ g^\B_\delta(\delta,\sigma)g^{\B'}_\delta(\delta,\sigma)\times {(\delta_{t_{3\B}, t_{3\B'}+1}+\delta_{t_{3\B'}, t_{3\B}+1})}{t_\B}.
\end{eqnarray*}}
\end{document}